# Rough subducting seafloor reduces interseismic coupling and mega-earthquake occurrence: insights from analogue models

Elenora van Rijsingen[1,2], Francesca Funiciello[1], Fabio Corbi[1], Serge Lallemand[2]

[1] Dep. of Sciences, Laboratory of Experimental Tectonics, "Roma Tre" University, Italy

[2] Géosciences Montpellier, CNRS, Montpellier University, France

Corresponding author : Elenora van Rijsingen (e.m.vanrijsingen@gmail.com)

## Key points

- We perform analogue models to investigate the effect of subduction interface roughness on megathrust earthquakes.
- Models with a very rough subduction interface are characterized by lower interface frictional strength and lower interseismic coupling.
- Ruptures in models with a very rough subduction interface are generally smaller in terms of rupture area, duration and mean displacement.

## Abstract

The roughness of the subduction interface is thought to influence seismogenic behavior in subduction zones, but a detailed understanding of how such roughness affects the state of stress along the subduction megathrust is still debated. Here, we use seismotectonic analogue models to investigate the effect of subduction interface roughness on seismicity in subduction zones. We compared analogue earthquake source parameters and slip distributions for two roughness endmembers. Models characterized by a very rough interface have lower interface frictional strength and lower interseismic coupling than models with a smooth interface. Overall, ruptures in the rough models have smaller rupture area, duration and mean displacement. Individual slip distributions indicate a segmentation of the subduction interface by the rough geometry. We propose that flexure of the overriding plate is one of the mechanisms that contribute to the heterogeneous strength distribution, responsible for the observed seismic behavior.

## Plain language summary

The largest and most destructive earthquakes on Earth occur along the plate contact in subduction zones, the region where an oceanic plate dives below another plate. The roughness of the downgoing plate, which is a result of the seafloor topography on that plate, is thought to play a role in the occurrence of large subduction earthquakes. With analogue models that include a 3D-printed seafloor, we test the effect of two types of seafloor roughness on the occurrence of earthquakes: a very rough vs. a very smooth seafloor. We observe that the rough seafloor geometry generally hinders the occurrence of large earthquakes along the subduction interface. This finding helps us to highlight where large future earthquakes are more likely to occur.

# 1. Introduction

The spatial occurrence of subduction megathrust earthquakes is thought to be influenced by the roughness of the subduction interface (e.g., Kelleher & McCann, 1976). This roughness mainly results from the size and distribution of topographic features on the seafloor, such as seamounts or ridges. Many studies have already addressed the influence of subducting topography on the spatial occurrence of megathrust earthquakes (e.g., Das & Watts 2009; Kopp 2013; Wang & Bilek 2014), but a detailed understanding of how this roughness affects the state of stress at the subduction interface, and therefore its seismogenic potential, is still debated.

By focusing on the spatial distribution of individual ruptures in nature, several studies have shown that a subducting seamount, ridge or fracture zone has acted as a barrier to rupture propagation (e.g., Das & Watts, 2009; Geersen et al., 2015; Henstock et al., 2016; Kodaira et al., 2000; Mochizuki et al., 2008; Robinson et al., 2006; Singh et al., 2011). In contrast, other theories suggest that a subducting feature may act as an asperity and therefore promote the occurrence of megathrust earthquakes instead (Bilek et al., 2003; Cloos, 1992; Husen et al., 2002; Scholz & Small, 1997). Recent studies have addressed this issue with a global approach and all converge to a model where smooth subduction interface is more prone to host large- to mega-earthquakes than a rough interface (Wang & Bilek 2014; Bassett & Watts 2015a; Lallemand et al. 2018; van Rijsingen et al. 2018).

Since the recurrence times for large megathrust earthquakes often exceed the natural record of ~100 years, models can be useful to study the process of subducting seafloor roughness over longer timescales and in a more systematic way. Among the first models of subducting seafloor topography were the sandbox experiments performed by Dominguez et al. (1998; 2000), which show a fracture network that develops in the overriding plate during single seamount subduction. Unfortunately, these models were not suitable for studying seismic behavior, since the material rheology does not allow stick-slip behavior. Since then, models addressing this topic that do include seismic behavior have been mainly numerical, such as the 2D sinusoidal fault models by Ritz & Pollard (2012), or the scale independent fault roughness models by Zielke et al. (2017), which both show that increasing fault roughness leads to smaller earthquake ruptures.

In this study, we address the problem by using seismotectonic analogue models (Corbi et al., 2013; Rosenau et al., 2017) that allow us to study the effect of fault roughness in a physically self-consistent, realistic and three-dimensional subduction setting, over the course of multiple seismic cycles. These models have been used before to study the synchronization of asperities on the interface by spatially varying the frictional properties from velocity weakening to velocity strengthening (Corbi et al., 2017a). Here we keep the frictional properties constant, but instead we introduce a 3D-printed geometry that represents the subduction interface. By reproducing scales of roughness that are in line with large topographic features observed in nature, we test if and how a rough interface influences the size and spatial distribution of megathrust earthquakes. Instead of focusing on a single seamount, we aim to look at a broader scale, allowing comparison with natural subduction zones that are characterized by a very rough- (e.g., Mariana) or very smooth (e.g., Kuril) subducting seafloor. The analogue setup allows us to only focus on the role of roughness of the interface, while keeping all other subduction parameters constant.

## 2. Methods

### 2.1. 3D-printing seafloor roughness

For studying the effect of subduction interface roughness on the occurrence of megathrust earthquakes, we use a 3D-printer to create two endmember-type subduction interfaces: a planar vs. a very rough interface (Figure 1a & b). Both interfaces include an isotropic, small scale roughness to ensure stick-slip behavior. It is characterized by peak amplitudes of 0.8 mm and a wavelength of 1 mm and shows stick-slip frictional behavior with amplitudes and periods comparable with the previously used sandpaper by Corbi et al. (2011, 2013, 2017a, 2017b). In addition to this small scale roughness, the rough interface is made up of a larger scale roughness, consisting of equally sized and homogenously distributed seamounts with amplitudes of 6.28 mm (4 km in nature) and a period of 94 mm (60 km in nature). These sizes are equivalent to large seamounts in nature, such as the Louisville Seamount Chain at the Tonga-Kermadec trench (Scholz & Small, 1997), or the Joban Seamount Chain in the Japan trench (Lallemand et al., 1989; Mochizuki et al., 2008).

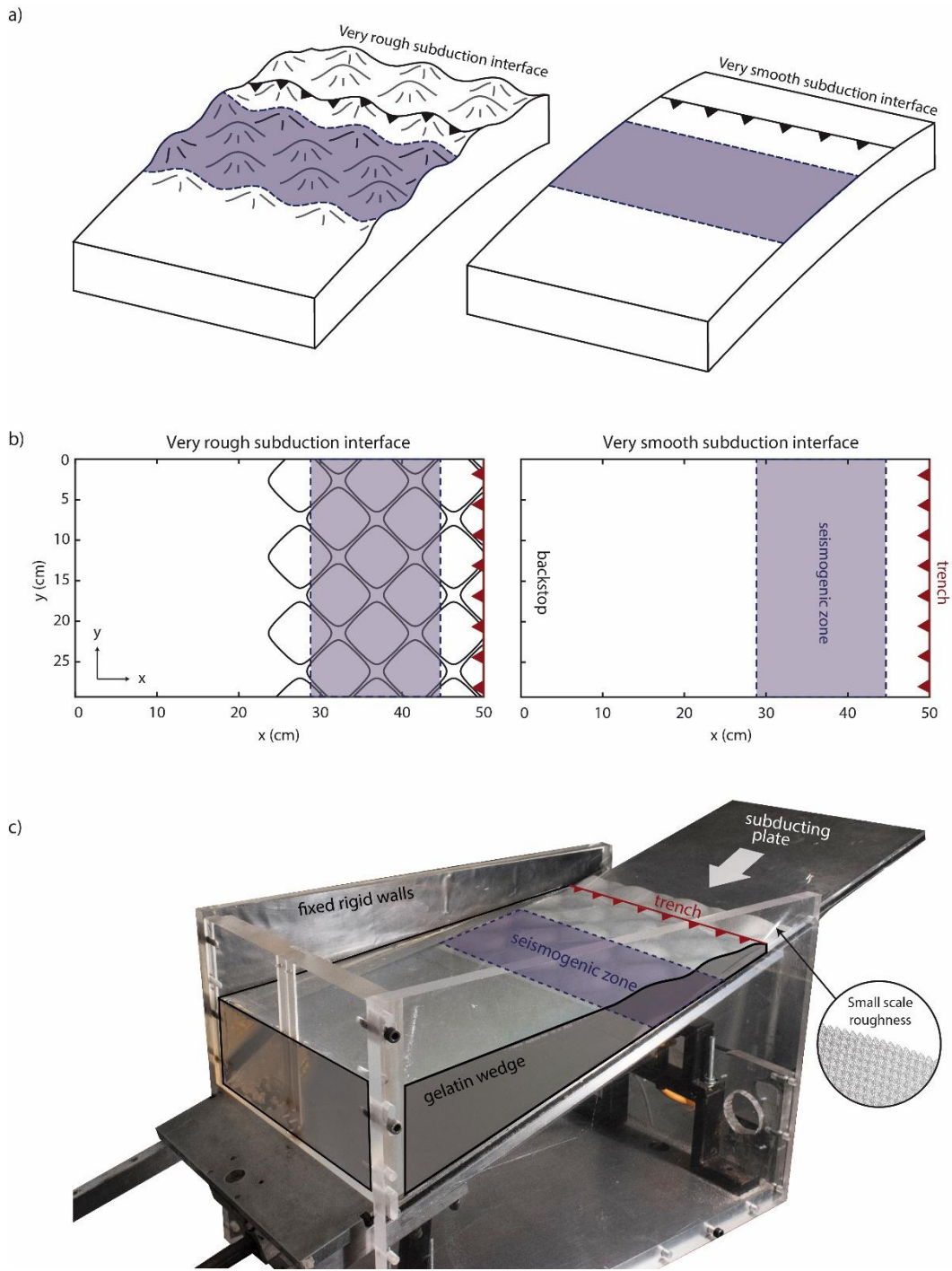

**Figure 1.** Model setup with two endmember interfaces. a) Cartoon illustrating the rough and smooth endmembers. b) Schematic representation (top view) of the experimental setup. The rough interface (black squares), the seismogenic zone (blue shaded rectangle) and the trench (red triangles) are indicated. c) Photograph of the experimental apparatus (oblique view). The inset shows a zoom on the small-scale roughness responsible for stick-slip behavior, that is superimposed on the large scale roughness.

## 2.2. Model setup and monitoring

The 3D-printed subduction interfaces are attached to a rigid, 10° dipping plate that represents the shallow portion of the subducting slab (Figure 1c). With a velocity of 0.1 mm/s, it underthrusts a gelatin wedge (2.5 wt% Pigskin; Di Giuseppe et al., 2009), the analogue of the overriding plate (for more details on the scaling of the gelatin wedge and the model setup with nature see the supporting information). A fixed plastic sheet with a window that indicates the seismogenic zone is placed between the downgoing plate and the gelatin wedge. Areas up- and downdip of the seismogenic zone will behave in a creeping manner, due to the contact between the plastic sheet and the gelatin wedge, while the 3D printed interface arriving within the window has stick-slip characteristics. In total, eight experiments are performed, from which four with an (identical) rough interface (Rough A-D) and four with a smooth interface (Smooth A-D). All experiments are monitored from above with a video camera that records 7.5 frames per second for a duration of 20 min, allowing us to observe tens of seismic cycles within one experiment. Images are post-processed using particle image velocimetry PIV (PIVlab, Thielicke & Stamhuis, 2014) resulting in displacement data at the top of the gelatin wedge, from which source parameters, such as rupture area and duration, interseismic coupling, recurrence time and mean displacement are extracted (see supporting information for details).

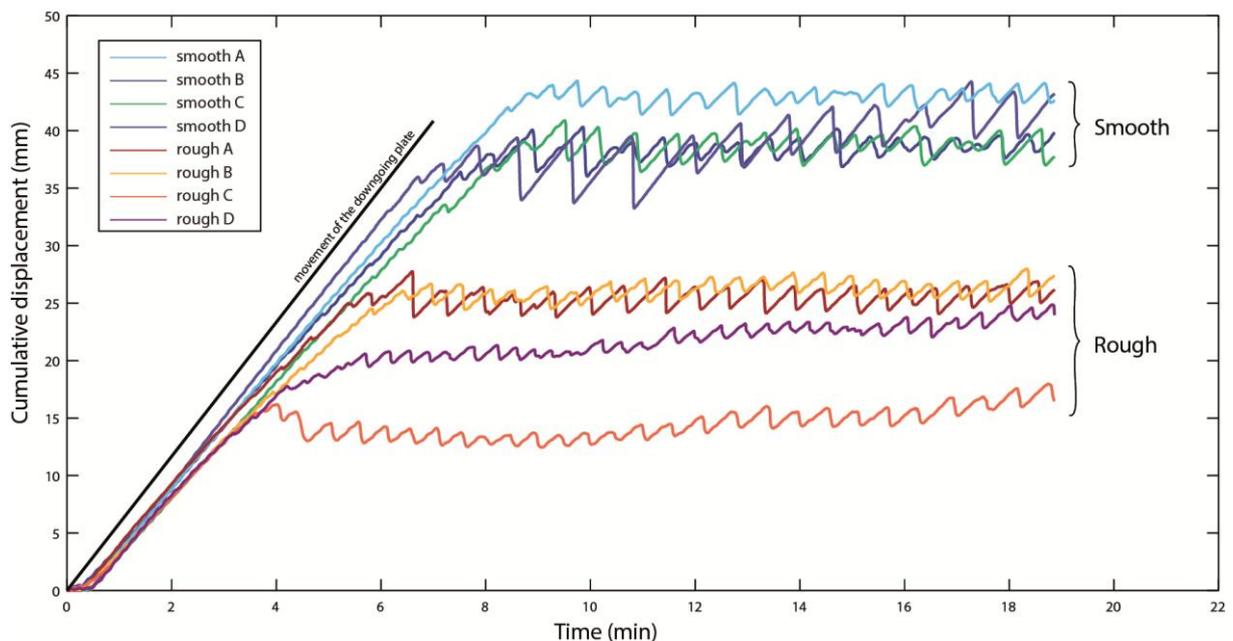

**Figure 2.** Cumulative displacement of one point centered above the seismogenic zone for each experiment. Each colored line represents one experiment, while the black line indicates the movement of the downgoing plate.

# 3. Results

## 3.1. General model behavior

All models go through an initial loading stage (lasting 4-8 min), during which the gelatin wedge gets elastically shortened and the trench slowly moves landward as the basal plate is underthrusted (Figure 2). At a certain strength threshold (after 5-10 % of shortening), the system starts to behave in a stick-slip manner, showing multiple seismic cycles characterized by a phase of landward loading (i.e., stick), followed by a quick release of stress and a seaward motion of (part of) the wedge (i.e., slip).

The initial loading coupling ranges from 67% (Rough D) to 82% (Smooth B) and the shortening threshold before stick-slip behavior lies between 15 mm (Rough C) and 43 mm (Smooth A). Looking at this shortening threshold, which is related to the frictional strength of the interface right before failure, we can clearly observe two groups: larger thresholds (i.e., larger interface strength) for all four smooth models, while the rough models show generally lower, and more variable shortening thresholds.

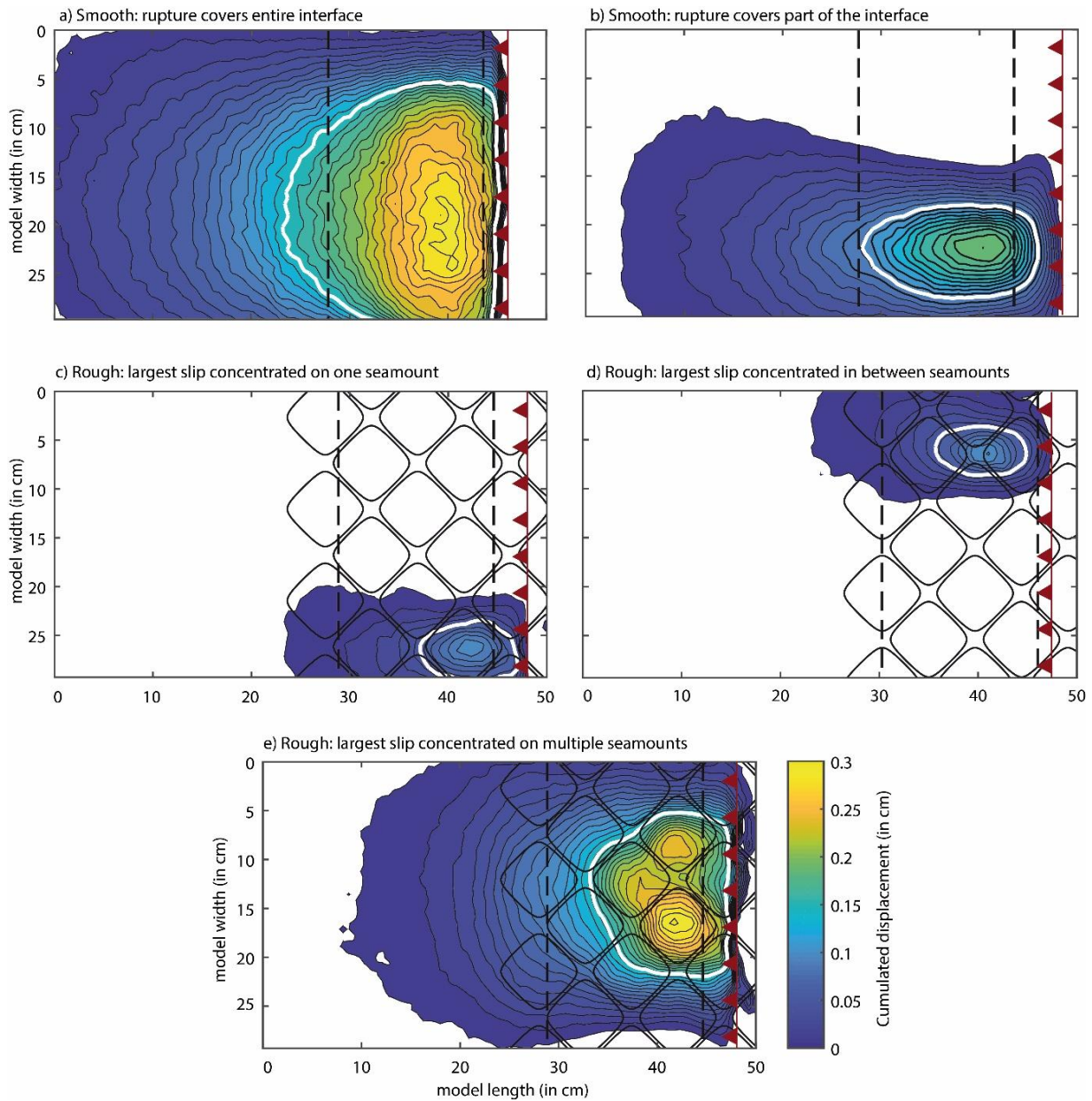

**Figure 3**. Cumulative displacement for representative coseismic events for the smooth- (a & b) and the rough models (c - e). Colors indicate cumulative slip in cm, following the colorbar on the bottom right. The white line represents the area where the maximum slip occurred (seismic asperity; 50% of maximum slip). The seismogenic zone is bounded by the dashed black lines and for the rough models, the seamount distribution is shown with solid black lines.

Figure 3 shows representative cumulative slip maps for both the smooth- and the rough models. For the smooth models, we can distinguish between ruptures that cover the entire interface (Figure 3a), and ruptures that cover only a third or half of the interface (Figure 3b). Ruptures in the rough models are generally smaller, often limited to only one or few seamounts. Looking at rupture evolution (see figure S1 in the supporting information), we observe crack-type ruptures, meaning that the nucleation region slips throughout the quake, expanding and then shrinking until the rupture stops (Marone & Richardson, 2006).

We analyzed all events in the rough models in terms of spatial distribution of the seismic asperity (white contours in Figures 3c-e). Ruptures were divided into three different categories: the maximum slip focused on a single seamount (33%), on multiple seamounts (14%), or mainly in between a group of seamounts (53%). These percentages show that only a small number of the events clearly ruptures multiple seamounts (Figure 3e), while most of the events have their maximum slip concentrated either on top of a single seamount (Figure 3c), or at a topographic low surrounded by seamounts (Figure 3d).

### 3.2. Source parameters: Rough vs. Smooth

Figure 4 shows violin plots for earthquake duration (a), recurrence time (b), interseismic coupling (c), mean slip (d), rupture area (e) and seismic asperity area (f). The smooth- and rough violin plots show the data distribution for all events in the smooth- (245 events) and the rough models (346 events). For earthquake duration, we observe a clear difference between the two endmembers. Rupture duration for the smooth models is generally much longer compared to the rough models. For the recurrence time we see similar distributions for both smooth and rough datasets. The interseismic coupling, which indicates how much movement of the downgoing plate is transferred to the overriding plate, is generally much higher for the smooth models than for the rough models. When looking at the slip parameters, we observe higher mean slip for events in the smooth models, ranging up to 2 mm, while mean slip for the rough models is smaller, with maxima at ~1 mm. Rupture areas in the smooth models appear to be slightly larger than the rough models, while the difference for the seismic asperity area is much clearer.

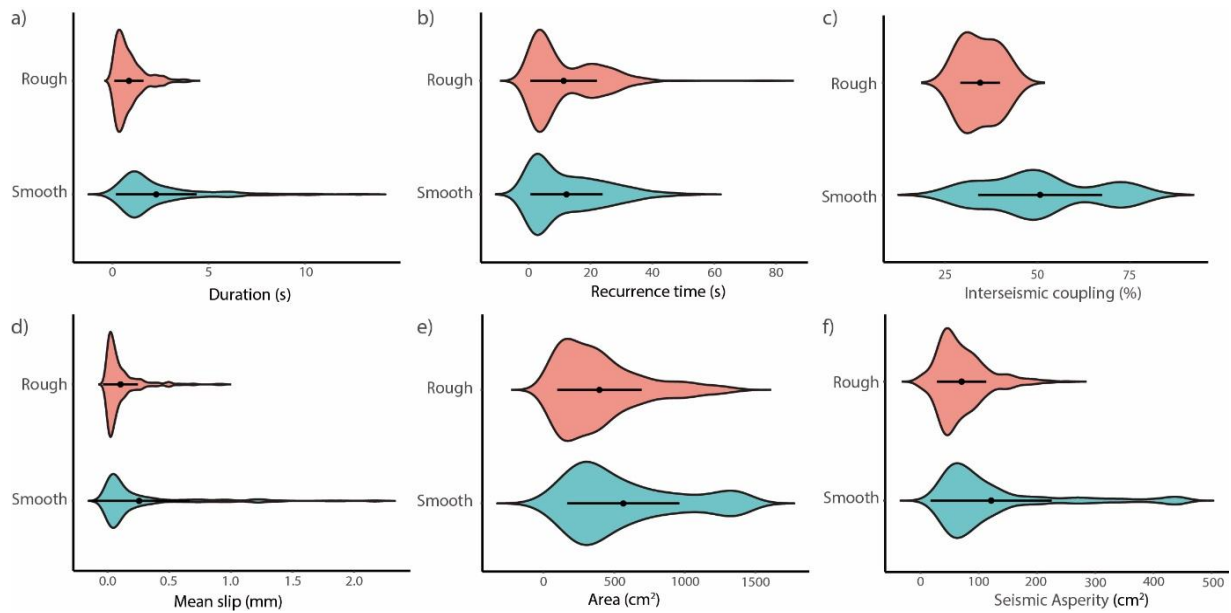

**Figure 4**. Violin plots of various source parameters for models with a smooth- (blue) and rough (red) subduction interface: rupture duration (a), recurrence time (b), interseismic coupling (c), mean slip (d), rupture area (e) and seismic asperity area (f). The curved lines show the distribution of the data, while the black dot and bar represent the mean and standard deviation, respectively.

## 4. Discussion

In this work, we study the effect of subduction interface roughness on the occurrence of megathrust earthquakes by using analogue models. We observe that models with a rough interface have lower interface frictional strength with respect to smooth models, meaning that they can store less elastic energy before failing (Figure 2). The finding that both initial coupling (during the strain accumulation phase) and interseismic coupling are also lower for rough models is therefore not surprising, since the ability for the interface to store elastic energy and the coupling between both plates are inherently related. From the source parameters, we observe a general trend for smaller earthquakes in the rough models, both in terms of rupture area and duration, as well as mean slip and seismic asperity area. The higher mean slip and larger seismic asperities in the smooth models indicate that the slip is more homogeneously distributed within the rupture area. The slip patterns for the earthquakes in the rough experiments confirm this, showing that the interface geometry seems to have a segmenting effect on the ruptures (Figure 3c-e).

## 4.1. Comparison with numerical modelling studies

The results from this study are in agreement with several numerical models that investigate the effect of fault roughness (Ritz & Pollard, 2012; Zielke et al., 2017) or seamount subduction (Yang et al., 2013) on the occurrence of earthquakes. Ritz & Pollard (2012) use a two-dimensional displacement discontinuity method to study the amount of slip and its distribution along an infinitely long sinusoidal interface in a homogeneous and isotropic elastic material. They find that where slip on a planar fault usually has an elliptical distribution, for the wavy fault interfaces the slip distributions are nonelliptical and reflect the sinusoidal geometry, something we see in our models as well (section 3.1). They also show that the mean slip decreases as the geometrical irregularity of the fault increases. Zielke et al. (2017) investigated the effect of fault surface roughness on the slip distribution and moment release by using large scale numerical simulations. They show that smooth faults have higher seismic moment releases, and therefore larger earthquakes than rougher faults. Yang et al. (2013) use 2D slip-weakening dynamic rupture simulations to investigate how a single geometrical high influences coseismic rupture propagation. They show that a seamount is more likely to act as a barrier for larger seamount height-to-width ratios, shorter seamount-to-nucleation distances and when normal stress at the seamount is increased with respect to the surroundings (as suggested by Kodaira et al., 2000). However, the additional normal stress required to stop rupture propagation decreases as the seamount height-to-width ratio increases.

## 4.2. Comparison with natural observations

Our model results are also in agreement with natural data. Global studies have shown that large earthquakes ($M_W \geq 7.5$) preferably occur along a smooth subduction interface and that rough subducting seafloor is associated with lower seismic coupling and a creep-like behavior (Bassett & Watts, 2015; Lallemand et al., 2018; van Rijsingen et al., 2018; Wang & Bilek, 2014). This trend can be illustrated by comparing two endmember subduction zones, the Izu-Bonin-Mariana trench versus the Japan-Kuril-Kamchatka trench. Following the classification of Lallemand et al. (2018), which is based on the seafloor characteristics prior-to-subduction, Izu-Bonin-Mariana is considered almost entirely rough, while Japan-Kuril-Kamchatka is dominantly smooth. This difference is also reflected in the occurrence of $M_W \geq$

7.5 megathrust events over the past ~100 years (van Rijsingen et al., 2018), since 96% of the Kamchatka-Kuril trench length has ruptured (among which 10 $M_W \geq 8.0$ ruptures), while no $M_W \geq 7.5$ events have occurred in the Izu-Bonin-Mariana trench. While the difference between rough and smooth seafloor seems clear on a global scale, different trends can be observed locally. In some places, geodetic measurements above subducting features show a local increase in interplate coupling (Collot et al., 2017; Kyriakopoulos & Newman, 2016), while in other places subducting topography is thought to cause a local decrease in coupling (Geersen et al., 2015; Marcaillou et al., 2016; Mochizuki et al., 2008; Singh et al., 2011). These local variations, but also the short natural record (i.e., ~100 years), make it challenging to come up with a mechanism that correctly explains the effect of subducting seafloor roughness on the occurrence of megathrust earthquakes.

## 4.3. Scenarios for subducting seafloor roughness

From the existing literature, two scenarios describing the effect of subducting relief on the coupling and seismogenic behavior in subduction zones can be considered. Scholz & Small (1997) argue that a local increase in normal stress above a subducting seamount would result in a higher coupling and therefore promote an asperity-like behavior (sensu Lay & Kanamori, 1981). Wang and Bilek (2011; 2014) however, suggest a decrease in coupling where topographic highs are subducting, due to a fracture network that is thought to develop around the features. Through this network of small faults, stresses along the interface are released, promoting a creep-like behavior rather than strain accumulation. This can explain the observed barrier-effect of subducting features (e.g., Geersen et al., 2015; Mochizuki et al., 2008; Singh et al., 2011), the decrease in seismic coupling for rough subduction zones (Lallemand et al., 2018), and the observation that large earthquakes preferably occur along a smooth megathrust (van Rijsingen et al., 2018). Also fluids that are delivered to the subduction interface may play a role in this, since fluid overpressures reduce the effective friction at the subduction interface, therefore reducing the amount of elastic strain accumulation (Bangs et al., 2006; Lallemand et al., 1994).

As in most numerical models covering this topic, the rheology of our overriding plate analogue (i.e., the gelatin wedge) is mainly elastic, and hence does not allow off-fault plastic deformation due to subducting topography. Therefore, one may expect results that are more in line with the scenario proposed by Scholz and Small (1997) (i.e., an increase in coupling when a topographic feature subducts). However, we observe a much lower interseismic coupling and interface strength for the rough models than for the smooth models (section 3.1 and 3.2). Therefore, even without off-fault plastic deformation, our results are still more in agreement with what is proposed by Wang and Bilek (2011, 2014): lower coupling and fewer large earthquakes when rough seafloor subducts. To explain our results, we propose that in our rough models, tensional stresses as a result of the flexure of the overriding plate create a very heterogeneous strength distribution and therefore segment the subducting interface (Figure 5.6).

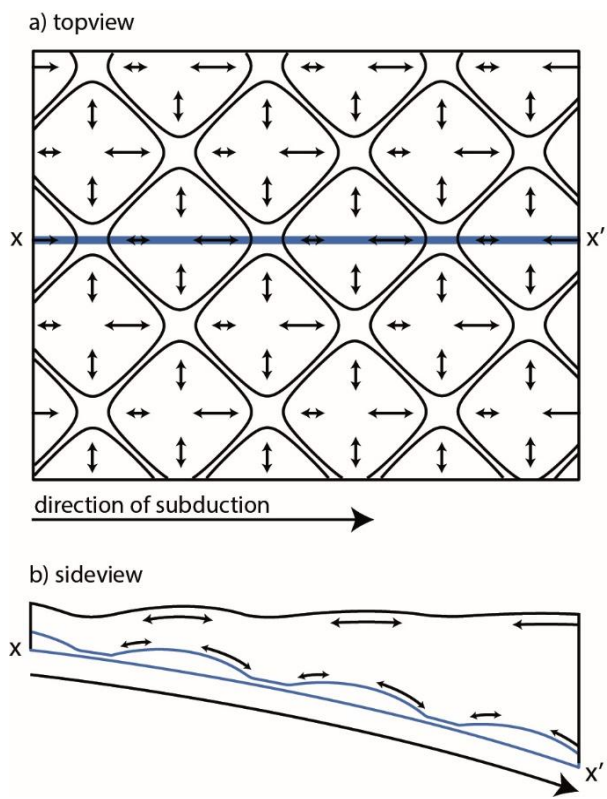

**Figure 5.** Schematic sketch of flexure of the overriding plate (highlighted by the black arrows) due to the rough geometry of the interface, topview (a) and sideview (b). Due to the dipping subducting plate, the flexure in the wedge will be larger along leading flanks of the seamounts. The tensional forces resulting from the flexure superimpose on the global wedge contraction (not depicted above) as basal shear undergoes.

Even though many ruptures in the rough models are relatively small, in some cases several seamount segments synchronize, creating a larger rupture with higher slip values. The segmentation of the interface due to the heterogeneous strength distribution might therefore make it more difficult for ruptures to propagate, but not impossible. In other words, the rough interface in our models hinders rupture propagation significantly, but does not prevent large ruptures to occur entirely (Nielsen & Knopoff, 1998). The more enhanced decoupling-effect of subduction interface roughness that is often observed in nature (e.g., Geersen et al., 2015; Mochizuki et al., 2008; Singh et al., 2011), could therefore be a combined effect of plastic deformation of the overriding plate (following the scenario of Wang and Bilek, 2011, 2014), the presence of fluids along the interface, as well as flexure of the overriding plate as proposed in this study, since they all contribute to a more heterogeneous strength distribution and therefore a segmentation of the seismogenic zone.

## 5. Conclusions

In this study we investigate the effect of subduction interface roughness on seismogenic behavior in subduction zones by using seismotectonic analogue models. We compared rupture source parameters and slip distributions for two roughness endmembers. We observe that models characterized by a very rough interface have lower interface frictional strength and lower interseismic coupling than models with a very smooth interface. In addition, ruptures in the rough models are smaller in terms of rupture area and duration, as well as mean displacement and seismic asperity. Their slip distributions and rupture evolution clearly reflect the segmenting effect of the rough interface geometry.

## Acknowledgements

We thank two anonymous reviewers for their detailed and constructive comments, as well as Silvia Brizzi, Joost van den Broek and Ajay Kumar for their help in the lab. All experimental data from this study are available through the EPOS data repository. This project has received funding from the European Union's Horizon 2020 research and innovation program under the Marie Sklodowska – Curie grant agreement 642029 – ITN CREEP. The grant provided to the Department of Science, Roma Tre University (MIUR – ITALY Dipartimenti di Eccelenza, ARTICOLO 1, COMMI 314 – 337 LEGGE 232/2016) is gratefully acknowledged.

Supporting Information for

**Rough subducting seafloor reduces interseismic coupling and mega-earthquake occurrence: Insights from analogue models**

E. van Rijsingen[1,2], F. Funiciello[1], F. Corbi[1], S. Lallemand[2]

[1] Dep. of Sciences, Laboratory of Experimental Tectonics, "Roma Tre" University, Italy

[2] Géosciences Montpellier, CNRS, Montpellier University, France

**Contents of this file**

Text S1 to S3

Table S1

Figure S1

**Introduction**

The supporting information contains additional information on the 3D-printing and modelling setup (text S1), about scaling of the analogue material and the experimental setup (text S2) and how the source parameters are extracted from the experiments (text S3). Text S1 is accompanied by Table S1, which contains details about the two scales of roughness that are included within the 3D-printed seafloor. Finally, Figure S1 shows the incremental and cumulative rupture evolution of the slip distribution displayed in Figure 3e.

**Text S1. Additional details on the 3D-printing and modelling setup**

To 3D-print the two types of subduction interfaces, a FlashForge Creator Pro 3D-printer is used, with PLA filament as printing material. With a Matlab algorithm, the 3D seafloor is designed, by stacking multiple 2D sinusoidal functions that are converted to positive values, only to make sure that the 3D-printed seafloor can be easily secured onto the downgoing plate of the analogue setup. Two scales of roughness are printed, a small scale roughness to ensure stick-slip behavior and a large scale roughness that represents a very rough seafloor characterized by large seamounts (Table S1.).

The 3D-printed subduction interfaces are attached to a rigid, 10° dipping rigid plate that represents the shallow portion of the subduction slab. It underthrusts a gelatin wedge that represents the overriding plate. A plastic sheet with a window of 34 x 16 $cm^2$ that indicates the seismogenic zone (216 x 102 $km^2$ in nature), is placed between the downgoing plate and the gelatin wedge. It follows the shape of the underlying topography, but stays in place during the experiment, meaning that the seismogenic zone (i.e., the window in the plastic sheet) will remain in a fixed place, while the downgoing plate passes below. Areas up- and downdip of the seismogenic zone will behave in a velocity strengthening way, due to the contact between the plastic sheet and the gelatin wedge, while the 3D-printed interface arriving within the window has velocity weakening characteristics.

|  | Amplitude A (mm) | Wavelength λ (mm) | Natural equivalent | Models |
|---|---|---|---|---|
| **Small scale roughness** | 0.8 | 1 | Provides frictional properties for stick-slip behavior | Both smooth and rough |
| **Large scale roughness** | 6.28 | 94 | Very rough seafloor characterized by large seamounts with heights of ± 4 km and widths of ± 60 km. | Rough only |

**Table S1.** Roughness properties used for the 3D printed subduction interfaces.

**Text S2. Scaling of experimental setup**

**S2.1. Scaling principle**

The analogue experiments are scaled to nature based on the principals of geometric, kinematic, dynamic, and rheological similarity (e.g., Hubbert, 1937). Each important physical dimension (i.e., length, time, and weight) is scaled to nature with a constant scaling factor (*). This is a dimensionless number that represents the ratio between model (M) and nature (N). Scaling factors for length (L*), density (ρ*) and viscosity (η*) are determined independently based on representative natural values, while scaling factors for stress (σ*) and time (T*) are derived from the other scaling factors. For time, two different scaling factors are used due to the very small experimental interseismic/coseismic time ratio: an interseismic- and a coseismic scaling factor (i.e., $T_i^*$ and $T_c^*$, respectively; Rosenau et al., 2009).

**S2.2. Scaling of analogue material**

The gelatin wedge used in the seismotectonic models is made out of 2.5 wt% Pig Skin gelatin. Di Giuseppe et al. (2009) explored the use of gelatin for analogue modelling, in order to have a single analogue material that can reproduce the complex rheological behavior of rocks. In their gel-state (i.e., solid-like behavior), gelatins show a visco-elasto-brittle rheology, while having a viscous rheology in their sol-state (i.e., fluid-like behavior). By varying gelatin composition, concentration, temperature, ageing and applied strain rate, Giuseppe et al. (2009) found that pig skin 2.5 wt% at 10° has the best rheological properties to serve as an analogue of the earth's crust, with a shear modulus (G) of $1 \times 10^3$ to $1 \times 10^4$ Pa, a viscosity (η) of $3 \times 10^5$ Pa s and a density (ρ) of 1000 kg/m$^3$.

**S2.3. Scaling principles applied to the experimental setup**

The experimental setup is scaled with a length scaling factor L* of $1.57 \times 10^{-6}$ (i.e., 1 cm in the model corresponds to 6.4 km in nature). From the density of the Pig Skin 2.5 wt% (i.e., 1000 kg/m$^3$), we can obtain density scaling factor ρ* of $3.45 \times 10^{-1}$ (assuming a natural density of 2900 kg/m$^3$), and following the relation:

$$\sigma^* = \rho^* \cdot L^*$$

a stress scaling factor σ* of $5.42 \times 10^{-7}$ can be determined (i.e., 1 Pa in the model corresponds to 1.85 MPa in nature). The shear modulus G follows the same scaling factor and scales with natural values ranging from $1.85 \times 10^9 - 1.85 \times 10^{10}$ Pa. The coseismic scaling factor is determined by assuming an instantaneous elastic response with a constant Froude number (i.e., the ratio of a body's inertia to gravitational forces). This results in a coseismic time scaling factor:

$$T_C^* = \sqrt{L^*}$$

(i.e., 1 s in the model corresponds to approximately 800 s in nature). On interseismic time scales, inertia is negligible, and viscous behavior becomes more dominant. This results in an interseismic time scaling factor:

$$T_i^* = \eta^* / \sigma^*$$

where the viscosity scaling factor, based on a natural viscosity of $5 \times 10^{21}$ Pa s, is $6 \times 10^{-17}$. This results in an interseismic time scaling factor of $1.11 \times 10^{-10}$ (i.e., 1 s in the model corresponds to 286 years in nature).

**Text S3. Source parameters**

The experiments are recorded from above with a video camera, recording 7.5 frames per second. Images are post-processed with Particle Image Velocimetry (PIV), resulting in a 2D velocity field for each frame (i.e., each 0.133 second). By working with surface displacements, we implicitly assume that these displacements are representative for the displacements at the subduction interface. From the velocity field, the maximum velocity, $V_{Xmax}$ (in the x-direction) for each frame can be extracted and peak velocities corresponding to a coseismic event can be identified. To isolate each coseismic event, a threshold of 0.05 cm/s is used, based on a trade-off analysis that shows the sensitivity of the number of events with respect to different thresholds. In this way, each series of frames for which $V_{Xmax}$ continuously exceeds this threshold is identified as one coseismic event. From this information, duration of each event can be determined by calculating the difference between the first and last frame of each event,

$$D_{eq}(i) = lastframe(i) - firstframe(i)$$

Following this, the recurrence time is calculated as the number of frames in between two consecutive events (1 frame = 0.133 s),

$$T_{rec}(i) = firstframe(i + 1) - lastframe(i)$$

The velocity of each cell in the 2D velocity field can be converted into displacement and eventually cumulative displacement over the course of one coseismic event. From the cumulative displacement map, the mean and maximum values can be extracted, which represent the mean- and maximum displacement during each event, respectively. By selecting all the cells that have exceeded the velocity threshold during the event, the total rupture area is calculated (the area of one cell is approximately 0.3 cm²). The seismic asperity area is calculated by selected all the cells that experienced ≥ 50% of the maximum displacement.

To calculate the interseismic coupling, the average surface displacement of the wedge towards the backstop (i.e., landward) in between two consecutive events is calculated as a ratio of the horizontal component of the displacement of the downgoing plate during the same time interval. Then, the average and standard deviation for all interseismic coupling values for both the rough- and smooth experiments are calculated.

The initial loading coupling (i.e., before initiation of the stick-slip behavior) is calculated in a similar manner, i.e. calculating the compression of the wedge towards the backstop as a ratio of the displacement of the downgoing plate for the first 4 minutes of each experiment.

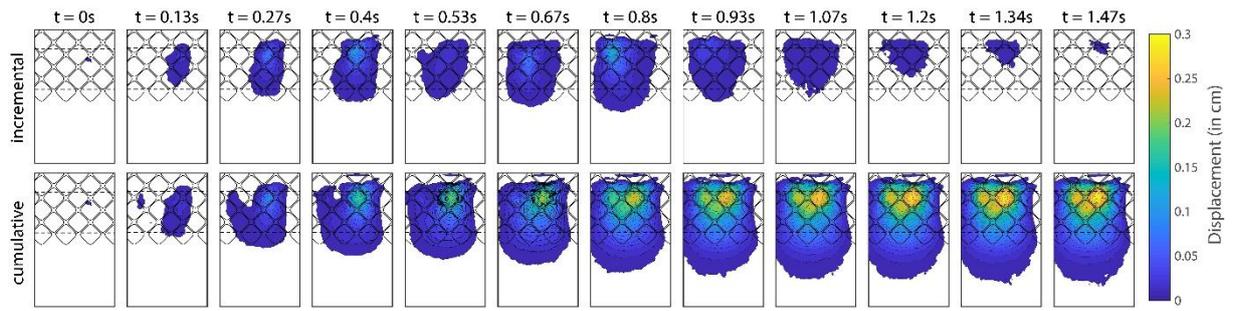

Figure S1. Incremental and cumulative slip distribution over time for an event rupturing multiple seamounts (Figure 3e). The seismogenic zone (dotted black lines) and the roughness pattern (solid black lines) are indicated in each sub-figure.